\documentclass [reprint,aip,jap]{revtex4-1}
\usepackage{graphics,graphicx}

\begin{document}

\title{Magnetic properties of Fe$_3$O$_4$ nanoparticles coated with oleic and dodecanoic acids}

\author{V. B. Barbeta}
\email{vbarbeta@fei.edu.br}
\affiliation{Departamento de F\'{i}sica, Centro Universit\'{a}rio da
FEI, Av. Humberto de A. C. Branco 3972, 09850-901, S. B. Campo,
SP, Brazil}

\author{R. F. Jardim}
\author{P. K. Kiyohara}
\affiliation{Instituto de F\'{i}sica,
Universidade de S\~{a}o Paulo, CP 66318, 05315-970, S\~{a}o Paulo,
SP, Brazil}

\author{F. B. Effenberger}
\author{L. M. Rossi}
\affiliation{Instituto de Qu\'{i}mica, Universidade de
S\~{a}o Paulo, 05508-000, S\~{a}o Paulo,
SP, Brazil}

\begin{abstract}
Magnetic nanoparticles (NP) of magnetite (Fe$_{3}$O$_{4}$) coated with oleic (OA)
and dodecanoic acids (DA) were synthesized and investigated through Transmission Electron Microscopy (TEM),
magnetization $M$, and \textit{ac} magnetic susceptibility measurements.  The OA coated samples were produced with different magnetic
concentrations (78, 76, and 65\%) and the DA sample with 63\% of Fe$_3$O$_4$. Images from TEM indicate
that the NP have a nearly spherical geometry and mean diameter $\sim$ 5.5 nm.
Magnetization measurements, performed in zero field cooled (ZFC)
and field cooled (FC) processes under different external magnetic fields $H$,
exhibited a maximum at a given temperature $T_{\rm B}$ in the ZFC curves, which depends
on the NP coating (OA or DA), magnetite concentration, and $H$. 
The temperature $T_{\rm B}$ decreases monotonically with increasing $H$
and, for a given $H$, the increase in the magnetite concentration results in an increase of $T_{\rm B}$.
The observed behavior is related to the dipolar interaction (DI)
between NP which seems to be an important mechanism in all samples studied. This is
supported by the results of the \textit{ac} magnetic susceptibility $\chi_{ac}$ measurements, where the temperature in which
$\chi'$ peaks for different frequencies follows the Vogel-Fulcher model, a feature
commonly found in systems with dipolar interactions. Curves of $H$ vs. $T_{\rm B}$/$T_{\rm B}$($H$=0) for 
samples with different coatings and magnetite concentrations collapse into a universal curve,
indicating that the qualitative magnetic behavior of the samples may be described by the NP themselves,
instead of the coating or the strength of the dipolar interaction. Below $T_{\rm B}$, $M$ vs. $H$ curves
show a coercive field ($H_{\rm C}$) that increases monotonically with decreasing temperature.
The saturation magnetization ($M_{\rm S}$) follows the Bloch's law and values of $M_{\rm S}$ at room temperature as high as 78 emu/g were estimated, a result
corresponding to $\sim$ 80\% of the bulk value.  The overlap of $M$/$M_{\rm S}$ vs. $H$/$T$ curves
for a given sample and the low $H_{\rm C}$ at high temperatures suggest superparamagnetic behavior in all samples studied.
The overlap of $M$/$M_{\rm S}$ vs. $H$ curves at constant temperature
for different samples indicates that the NP magnetization behavior is preserved,
independently of the coating and magnetite concentration.
\end{abstract}

\maketitle

\section{Introduction}

Iron oxides have many different technological applications, ranging from coloring 
of glasses to magnetic recording.  In the latter, the increase of data density brings 
new challenges due to the decrease of the size of magnetic particles down to a nanometric 
scale which make systems unstable by the occurrence of superparamagnetism (SPM).
In addition to this, the dipolar interaction (DI) between particles plays an 
important role in the general magnetic properties, being responsible 
for the gradual signal degradation with time.\cite{AZE1} Although many researches
have focused in iron oxides nanoparticles (NP) for practical applications,
there are still some controversial issues like the role played by the DI in 
the magnetic field dependence of the magnetization.\cite{AZE1} Among iron oxides, 
magnetite (Fe$_{3}$O$_{4}$) is a material that displays interesting magnetic properties, 
mostly when the particles are in the nanometric scale.

Bulk magnetite is a ferrimagnetic compound with Curie temperature close to 860 K.\cite{NEE1} 
The oxygen anions form a face-centered cubic lattice with Fe$^{2+}$ and Fe$^{3+}$ cations in 
interstitial sites. The tetrahedral ({\it A}) sites are occupied by Fe$^{3+}$, and the octahedral 
({\it B}) sites are randomly occupied by both Fe$^{3+}$ and Fe$^{2+}$, resulting in an inverse 
spinel structure.\cite{HAM1} There are a number of applications for Fe$_{3}$O$_{4}$ NP such as for 
recovering Ru catalysts from liquid-phase oxidation and hydrogenation reactions,\cite{JAC1} as a 
contrast agent for magnetic resonance imaging (MRI) in biological tissues, and for hyperthermia in 
experiments of cancer treatments,\cite{TAR1} among others.

Clustering of magnetic NP can be avoided by coating, where a layer of surfactant acts in changing 
the interaction between NP by altering the strength of the dipolar interaction.  Different types 
of coating materials have been used for this purpose including silica and oleic acid. It is largely 
accepted that the oleic acid creates a non-magnetic and superficial single-layer surrounding the NP, 
thus reducing the magnetic interaction.\cite{ZHA1}

Another feature related to the reduction of the dimension of magnetic granules is the decrease of 
the saturation magnetization ($M_{\rm S}$), which is smaller in NP than its  corresponding bulk value. 
The reduction of $M_{\rm S}$ in NP is a controversial issue with arguments in favor of finite size 
effects and surface spin disorder.\cite{MAR1} These effects are much more pronounced in ferrimagnetic 
systems like magnetite.  In this case, the superexchange interaction between Fe ions is mediated 
through O$^{2-}$ ions,\cite{BER1} and incomplete coordination at the surface and/or oxygen vacancies 
are believed to be responsible for a surface spin disorder. This makes the total magnetization of the 
NP much smaller than the bulk value.  Therefore, $M_{\rm S}$ usually assumes lower values due to the 
increase of the surface/volume ratio.\cite{CAI1} The higher is the saturation magnetization, the more 
important is the material for practical applications.  Therefore, it is important to understand how the 
saturation magnetization can be increased in a system of NP to make it valuable for practical applications.  
Previous studies indicated that covering magnetite NP with oleic acid (OA) results in higher values 
of $M_{\rm S}$ due to the lowering of the surface magnetic disorder,\cite{GUA1} but the overall effect 
of the surfactant coating in the NP magnetic behavior is still an open question.

In this work we have investigated a series of magnetite NP prepared with two different coatings: 
oleic acid (OA samples) with different concentrations of magnetite (O65-65\%, O76-76\%, O78-78\%), 
and dodecanoic acid (D63-63\%). The samples were magnetically characterized through both \textit{ac} 
and \textit{dc} magnetic susceptibility measurements in order to gain further information regarding 
the influence of changing both the coating and the magnetic concentration in their overall magnetic behavior.

\section{Experimental Procedure}

The synthesis of the OA coated NP was performed through a modified protocol
for decomposition of an Iron(III) precursor in a high-temperature solution
phase reaction, as described elsewhere.\cite{SUN1,SUN2}  In a typical preparation,
2 mmol of Fe(\textit{acac})$_3$ was dissolved in 6 mmol of OA, 4 mmol of oleylamine and
20 mL of phenyl ether, followed by addition of 10 mmol of 1,2-octanediol under
vigorous magnetic stirring and flow of N$_{2}$.  The final mixture was heated at
210 $^\circ$C and refluxed for 2 hours under N$_{2}$ atmosphere.  After the mixture was
cooled down to room temperature, the Iron oxide NP were precipitated by adding ethanol
and separated through centrifugation at 7000 rpm.  The process was repeated several
times until the supernatant solution became clear.  Then, the NP were dried under
vacuum and subjected to elemental analysis (CHN) in order to determine the amount
of magnetic material.  Samples containing 78, 76, and 65 \% of mass corresponding
to Iron oxide were prepared by adding or removing OA from the material prepared
according to the procedure described above.  
In order to remove OA from the surface of the NP, the samples were subjected to reflux in acetone under N$_{2}$ atmosphere for 2 hours.

The DA coated NP were obtained by using the same magnetite NP described above, after the OA excess removal from the sample. 6 mmol of
iron oxide NP previously obtained were then heated with 9 mmol of dodecanoic acid and 10 mL
of toluene as solvent at 80 $^\circ$C for 40 minutes. After the mixture was cooled down to
room temperature, the DA coated NP were precipitated by adding ethanol/acetone (1:1 v/v)
and separated via centrifugation at 7000 rpm.  The NP were dried under vacuum and subjected
to elemental analysis. By combining the two procedures described above, we have successfully 
synthesized samples comprised of \textit{the same magnetite NP} coated with both OA and DA and different 
concentrations of the magnetic material.

Transmission Electron Microscopy (TEM) was used to gain information
regarding the morphology of the NP. The TEM images were obtained 
on a Philips CM 200 microscope operating at an accelerating
voltage of 200 kV.  The samples for TEM observations were prepared by placing a drop of a
toluene solution containing the NP in a carbon-coated copper grid. The histograms of the
NP size distribution, assuming spherical shape, were obtained from measurements of about
600 particles found in arbitrarily chosen areas of enlarged micrographs
of different regions of the Cu grid.

All magnetization $M(T,H)$ and \textit{ac} magnetic susceptibility $\chi_{ac} = \chi ' + i\chi ''$ measurements  
were performed in a Quantum Design MPMS SQUID magnetometer
for the monodispersed samples in powder form.  The $\chi_{ac}$ vs. $T$ curves were obtained using
an excitation field of $\sim$ 1 Oe, in a wide range of frequencies, and under zero external \textit{dc} magnetic field.
The $M$ vs. $T$ measurements were performed under
both zero-field cooled (ZFC) and field cooled (FC) conditions. The ZFC cycle was performed
after heating up the sample up to 300 K, well above the blocking temperature, and cooling
down the sample to 5 K without applied magnetic field.  
After this step, a small measuring magnetic field $H$ was
applied, and the data were collected increasing the temperature from 5 to 300 K.  In order
to avoid remanent field from the superconducting coils, before cooling down the sample for
the ZFC run the superconducting magnet has been reset three times.  The FC measurements were
performed after the ZFC cycle, cooling down the sample and keeping the same external magnetic
field applied for the ZFC measurement.  After each FC cycle, the temperature was increased to
300 K, the magnet has been reset, and the sample cooled down to 5 K for successive measurements.

The $M$ vs. $H$ measurements were performed after cooling the samples down to a desired
temperature under zero external magnetic field.  Once the temperature was reached, $H$ was
cycled from 7 T to - 7 T, back and forth.  After collecting the data for each $M$ vs. $H$ curve,
the temperature was increased up to 300 K and the magnet reset three times,
avoiding any remanent field for the subsequent measurement.

\section{Results and Discussion}

\subsection{Transmission electron microscopy}

The results of TEM obtained for all samples indicated that the 
NP are uniform in shape and size and have a nearly spherical shape, as inferred from the image displayed in Figure 1.  
The diameter distribution was obtained by counting over than 600 NP
and the data were fitted to a lognormal distribution, resulting in
a mean diameter $d_{\rm T} \sim 5.5$ nm with a distribution
width $s_{\rm T} \sim$ 0.15, and standard deviation $\sigma_{\rm T}
\sim$ 0.7 nm.

\begin{figure} [htp]
\centering
\includegraphics [width=0.2\textwidth] {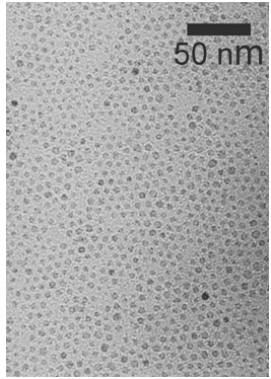}
\caption{\label{fig:epsart1} TEM image obtained for the sample O76.}
\end{figure}

The critical diameter $d_{\rm C}$ for monodomain formation in magnetic materials is given by $d_{\rm C}  \sim 18\frac{\sqrt{AK}}{\mu_{\rm 0}{M_{\rm S}}^2}$, where $A$ is the exchange constant, $K$ is the magnetic anisotropy, and $M_{\rm S}$ is the saturation magnetization. The critical diameter for the magnetite can be estimated by using the bulk values of $A$, $K$, and $M_{\rm S}$ ($A \sim$ 1.3x10$^{-11}$ J/m, $K \sim$ 1.35x10$^{4}$ J/m$^{3}$, and $M_{\rm S}$ $\sim$ 4.6x10$^{5}$ A/m), resulting in $d_{\rm C} \sim$ 28 nm. Therefore, the diameter of the magnetite nanoparticles studied here is well below 
its critical size, indicating that all particles, even the large ones, 
can be considered as single domain.

\subsection{Magnetization versus temperature}

The \textit{dc} magnetization curves $M$ vs. $T$ for all samples studied 
were found to exhibit similar features. Figure 2 shows typical
curves obtained for the sample O76 under FC and ZFC conditions and for
different measuring magnetic fields. A common feature of these curves
is the occurrence of a well defined maximum in ZFC curves at a
temperature $T_{\rm B}$, which is usually identified as the NP
blocking temperature.  As $H$ is increased, $T_{\rm B}$ shifts monotonically towards lower temperatures.

\begin{figure} [htp]
\includegraphics{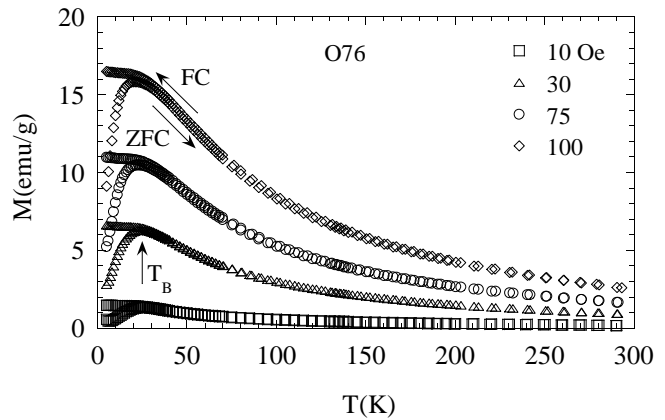}
\caption{\label{fig:epsart2} ZFC and FC curves for sample O76 under different
applied magnetic fields.}
\end{figure}

By taking pairs of ($T_{\rm B}$, $H$) from similar curves as those displayed in Figure 2, one
is able to build a $T_{\rm B}$ vs. $H$ phase diagram.  Such a phase diagram, for all
samples studied, is displayed in Figure 3. The inset shows $H$ as a function of the normalized
temperature $T_{\rm n}=T_{\rm B}$/$T_{\rm B}$($H$=0). From the results of Figure 3, some general features can be observed: (i) $T_{\rm B}$ is a monotonically decreasing
function of $H$; (ii) increasing the concentration of the magnetic material results 
in an increase of $T_{\rm B}$; (iii) all data follow a nearly universal curve when 
$T_{\rm B}$ is replaced by the normalized temperature $T_{\rm n}$=$T_{\rm B}$/$T_{\rm B}$($H$=0), as 
shown in the inset of the Figure 3.

\begin{figure} [htp]
\includegraphics{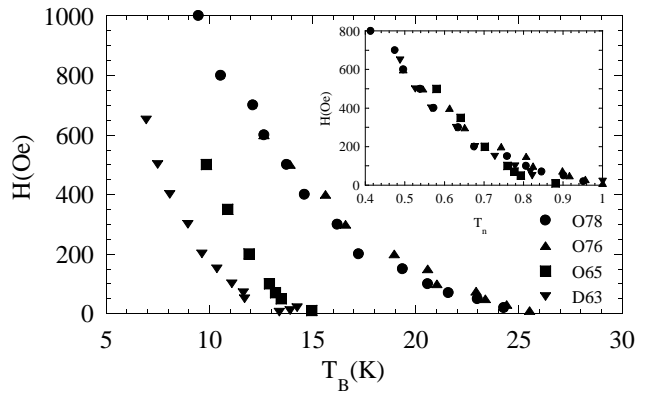}
\caption{\label{fig:epsart3} Curves of $H$ versus $T_{\rm B}$ for all samples studied.
The inset displays $H$ as a function of the normalized temperature $T_{\rm n}$=$T_{\rm B}$/$T_{\rm B}$($H$=0), as discussed in the text.}
\end{figure}

The systematic decrease of $T_{\rm B}$ with increasing $H$ is frequently seen in systems 
comprised of magnetic NP. Such a behavior is related to the presence of dipolar interaction 
between NP and usually found in systems where the concentration of the magnetic material is 
high.\cite{KAC1} This is the case of our samples, in which the concentration of the magnetic 
material reaches values as high as 78\%. On the other hand, the behavior of $T_{\rm B}$ 
with $H$ observed here is quite different from the one found in frozen ferrofluids of magnetite NP suspended in a
nonmagnetic solvent.\cite{LUO1} In such a system, where the magnetic concentration of NP is actually 
very low, the $T_{\rm B}$ vs. $H$ phase diagram displays a well defined reentrant behavior. 
There, $T_{\rm B}$ increases with increasing $H$ at low magnetic fields, passes through a maximum value, 
and decreases with increasing $H$ at high fields.  Thus, the dipolar interactions in our samples 
are important and responsible for the $T_{\rm B}$ vs. $H$ behavior displayed in Figure 3.

In fact, changes in the volume fraction of the magnetic material have their counterpart in the magnitude of $T_{\rm B}$($H$=0), 
obtained by extrapolating the $H$ vs. $T_{\rm B}$ curves to $H$= 0. We have found that increasing the concentration 
of the magnetic material results in a systematic increase of $T_{\rm B}$($H$=0), which ranged from $\sim 14$ K 
for the sample D63 to $\sim 25$ K for the sample O78. This result indicates that the increase in magnetic concentration
leads to a progressive increase in $T_{\rm B}$, due to the increase in the interparticle interactions.  Such an increase of $T_{\rm B}$ with increasing magnetic volume has been predicted in theoretical models,\cite{DOR1} as well as in Monte Carlo 
simulations,\cite{OTE1} where dipolar interactions between magnetic NP are considered.

We also mention that the universal behavior observed in the $H$ versus $T_{\rm n}$=$T_{\rm B}$/$T_{\rm B}$($H$=0) 
curves (see inset of Fig. 3) is a convincing manifestation that all samples are comprised of \textit{the same magnetite NP}. 
Under this circumstance, it seems that the only one effect related to the DI between NP is to shift $T_{\rm B}$ to higher values, while features 
of the individual NP are preserved. The relevance of the DI in establishing $T_{\rm B}$ 
is also supported by its value, which can be estimated for a system comprised of noninteracting NP. In this case, a rough estimate of the blocking temperature is obtained by using the expression $25k_{\rm B}T_{\rm B} \sim E$,
where $E = KV$, $K$ is the anisotropy constant ($K \sim$ 1.35x10$^{4}$ J/m$^{3}$ for the bulk magnetite), $k_{\rm B}$ is the
Boltzmann constant, and $V$ is the NP volume. Using the average diameter obtained in TEM images, 
the blocking temperature $T_{\rm B}$ in the
noninteracting limit was found to be $\sim$ 3 K, a value much lower than $T_{\rm B}$($H$=0) $\geq$ 14 K measured in all samples, further indicating the role played by the DI in our samples.

\subsection{ac magnetic susceptibility versus temperature}

The dynamic properties of the samples were also investigated by \textit{ac} magnetic susceptibility measurements.  
Typical data of the temperature dependence of both components of the $\chi_{ac}$ ($\chi '$ and $\chi ''$) for the sample O78 
are displayed in Figure 4. The data were taken in a large range of frequencies $f$, from 0.021 Hz up to 957 Hz, 
and under zero applied magnetic field. Some important features in these curves are: (\textit{i}) the 
occurrence of a frequency dependent rounded maximum at $T'_{\rm m}$ and $T''_{\rm m}$ on
both $\chi'$ and $\chi''$ components, respectively; (\textit{ii})
increasing $f$ results in a shift of both $T'_{\rm m}$ and
$T''_{\rm m}$ to higher temperatures; (\textit{iii}) $\chi'$ is essentially 
$f$ independent at high temperatures $T \gg T'_{\rm m}$,
indicating a superparamagnetic behavior of the Fe$_{3}$O$_{4}$ NP;\cite{Goya} and (\textit{iv})
a frequency dependent behavior of $\chi'$ for $T \leq T'_{\rm m}$,
further suggesting the blocking process of the Fe$_{3}$O$_{4}$ NP.\cite{DOR3} 

\begin{figure}[htp]
\centering
\includegraphics{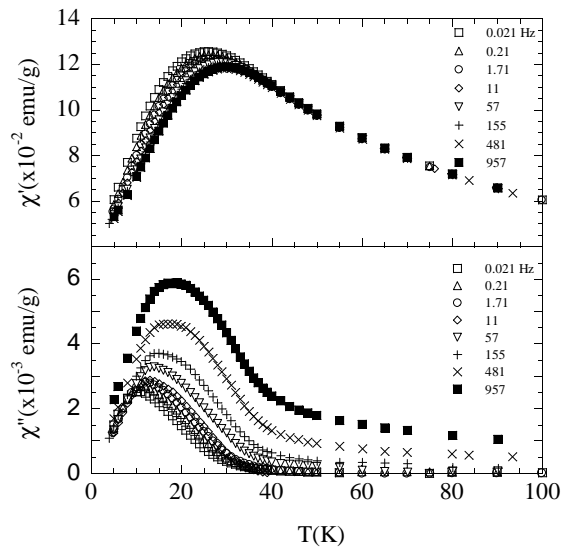}
\caption{\label{fig:epsart4} Curves of $\chi'$ and $\chi''$ vs. $T$ obtained for sample O78 using different 
frequencies of the excitation field.}
\end{figure}

We also mention that the magnitude of the $\chi ''$ component increases appreciably with 
increasing $f$, a feature commonly related to the presence of dipolar interactions between 
NP.\cite{JON} Thus, mostly based on the results described above, it is tempting to 
discuss whether dipolar interactions between NP are responsible for the overall behavior 
of the $\chi_{ac}$ data displayed in Figure 4. Such a discussion can be further 
explored by considering the inverse of the \textit{ac} frequency $\tau=1/f$ as a 
function of the temperature $T'_{\rm m}$ in which $\chi'$ peaks. A 
noninteracting system is believed to obey the N\'{e}el-Brown model, i.e., to 
follow an Arrhenius law $\tau=\tau_{0}exp(E/k_{\rm B}T)$,\cite{BAT1} where $\tau_{0}$ is
the characteristic relaxation time, $E$ is the anisotropy energy, and $k_{\rm B}$ is the
Boltzmann constant. The inset of Figure 5 shows the fitting to an Arrhenius law for the data of Fig. 4,
resulting in $\tau_{0} \sim 10^{-32}$ s, and $E \sim$ 1900 K. Although the 
fitting result is not poor, the obtained $\tau_{0}$ value is unphysically small and out 
of the expected range (roughly from $10^{-12}-10^{-9}$
s) as well as the anisotropy energy $E$ is exceptionally high when compared with $\sim$ 80 K,
estimated by using the bulk value of $K$ and $d \sim$ 5.5 nm. These combined results indicate that magnetic interactions between NP must 
be considered in any analysis performed in these samples.

On the other hand, it has been suggested that magnetic interactions between NP may result 
in an ordered state, with a spin-glass like behavior. In this case,
a divergence of the relaxation time $\tau$ is expected to occur at a glass-like transition temperature $T_{\rm g}$, 
according to the conventional critical slowing down $\tau=\tau_{0}(T/T_{\rm g}-1)^{-z\nu}$.\cite {DJU}
The critical exponent ${z\nu}$ is expected to range from $\sim$ 4 up to $\sim$ 12 in conventional spin-glasses,\cite {MYD} and
values like $11\pm3$ have been reported for NP.\cite {DJU} Fitting our data to this scale law resulted in a good fitting (not shown) with $\tau_{0} \sim 2.10^{-9}$ s, $T_{\rm g} \sim 22$ K but a very large value of $z\nu \sim 17$. The $\tau_{0}$ value is compatible with results 
observed in NP of $\gamma$-Fe$_2$O$_3$ \cite{DOR2} and Fe-C,\cite{HAN2} although the $z\nu$ values reported elsewhere 
were always much smaller ($z\nu \sim 7$ and $\sim 10$, respectively). Thus, based on this analysis, the existence of a critical behavior and the presence of a phase transition to a glass-like state can be disregarded in our samples.

Another approach to disclose the effect arising from the dipolar interactions between NP is to consider 
an extension of the noninteracting case (the N\'{e}el-Brown model) and similar to the above mentioned scale law, called the Vogel-Fulcher 
law, in which the relaxation time is given by $\tau=\tau_{0}exp[E/k_{\rm B}(T-T_{0})]$.  Such a law has been frequently used for describing the dynamic properties 
of systems in which the volume of the magnetic component is above a certain value and $T_{0}$ is a measure of the strength of 
the interparticle dipolar interaction.\cite{SUE} Thus, we have fitted our $\chi_{ac}$ 
data to the Vogel-Fulcher law and found an excellent agreement by using $T_0 \sim 17$ K, and $\tau_{0} \sim 10^{-13}$ s, 
as displayed in Figure 5. Such a result indicates that our $\chi_{ac}$ data are consistent with a system in which 
dipolar interactions between NP are responsible for its dynamic properties.

\begin{figure}[htp]
\includegraphics{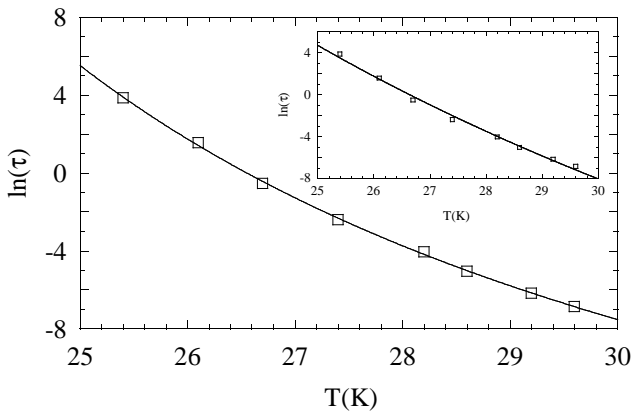}
\caption{\label{fig:epsart5} Fitting of the temperature dependence of ln$\tau$ to the Vogel-Fulcher function for the sample O78. The inset shows the fitting to
an Arrhenius law for the same sample.}
\end{figure}

We finally argue that in order to distinguish between a glass-type transition 
from superparamagnetic interacting NP, a model-independent empirical parameter $\phi=\Delta T_{\rm M}/(T_{\rm M} \Delta log \nu)$ with $T_{\rm M}$, estimated for 
$\nu = $ 50 Hz,\cite{DOR3} can be considered. The resulting $\phi \sim 0.032$ value is higher than
those observed in conventional spin-glass systems ($\phi \sim 0.005-0.015$), 
and slightly smaller than the ones found in coupled granules ($\phi \sim 0.05-0.13$).\cite{DOR4} 
On the other hand, the ratio $(T_{\rm g}-T_{0})/T_{\rm g} \sim 0.38$,
evaluated near $\nu = 10$ Hz and $\nu_{0} = 10^{13}$ Hz and with $T_{0}$ obtained from
the fit to the Vogel-Fulcher law, resulted in values higher than those observed for 
spin-glasses ($\sim 0.07-0.15$), but in line with the ones expected for weakly-coupled 
interacting NP ($\sim 0.25-1.0$).\cite{DOR4} This result reinforces the idea that
for the samples studied here there is no evidence for a phase transition to a glass-like phase, and
that $T_{0}$ represents a measure of the dipolar interaction.

\subsection{Magnetization versus magnetic field}

The $M$ vs $H$ curves for all samples studied were found to show similar behavior.
Figure 6 displays typical $M$ vs. $H$ curves obtained for sample O76 at two
selected temperatures: 5 and 300 K. The inset of Fig. 6 exhibits the $M$ vs. $H$ 
behavior at low magnetic fields. At low temperatures ($T < T_{\rm B}$) the 
system is in the blocked state and curves $M$ vs. $H$ exhibit 
both remanence ($M_{\rm R} \sim 5$ emu/g) and coercivity ($H_{\rm C}\sim 50$ Oe).
Our $M$ vs. $H$ data at several temperatures below $T_{\rm B}$ also indicated that 
$H_{\rm C}$ is symmetrical regarding the positive and negative values of 
the magnetic field, as well as $M_{\rm R}$, indicating that exchange bias are absent in our samples. 
We have also observed that at temperatures just above $T_{\rm B}$, 
both the remanence and the coercive field become negligible.

\begin{figure}[htp]
\centering
\includegraphics{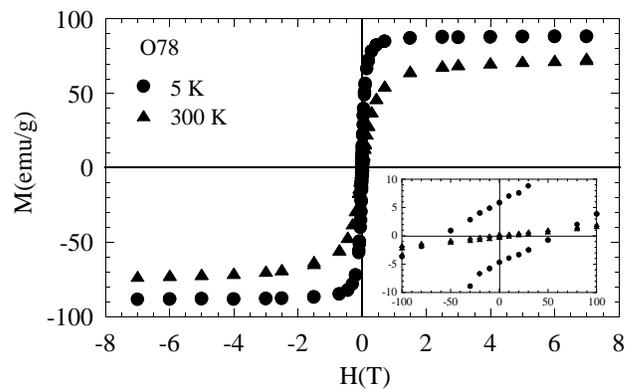}
\caption{\label{fig:epsart6} $M$ vs. $H$ curves for sample O76 at 5 and 300 K.
The inset shows an expanded view of the low magnetic field behavior for both temperatures.}
\end{figure}

\begin{figure}[htp]
\centering
\includegraphics{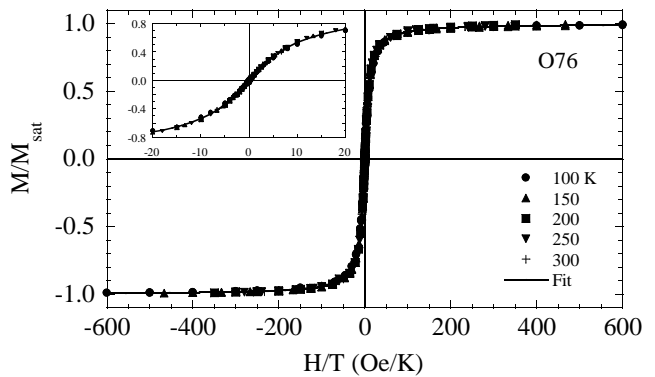}
\caption{\label{fig:epsart7}Curves of $M/M_{\rm S}$ vs. $H/T$ at several temperatures
$T > T_{\rm B}$, for sample O76.  The inset shows an expanded view of the region with low $H/T$.}
\end{figure}

All $M/M_{\rm S}$ vs. $H/T$ curves, where $M_{\rm S}$ is the experimental saturation magnetization
obtained at H = 7 T, measured at temperatures higher than $T_{\rm B}$ were found to overlap in a
common curve, as displayed in Figure 7 for the sample O76.  Such an overlap of the $M/M_{\rm S}$ vs. $H/T$ curves
in all samples along with the near-zero $H_{\rm C}$ clearly indicate the occurrence of 
superparamagnetism at temperatures higher than
$T_{\rm B}$ and below the bulk critical ferrimagnetic ordering temperature $T_{\rm C}$.

In the superparamagnetic phase, the collapsed $m$ vs. $h$ ($m = M/M_{\rm S}$, and $h = H/T$)
curves are well described by a Langevin function $L(x)$ weighted by a
lognormal magnetic moment distribution $f(\mu)$, as discussed elsewhere.\cite{FON1} Figure 7 shows 
such a fitting for sample O76 and table I displays the mean magnetic moment ($\mu_{\rm m}$), 
mean magnetic diameter ($d_{\rm M}$),
distribution width ($s_{\rm M}$), and standard deviation ($\sigma_{\rm M}$), obtained 
from the fitting procedure. Table I also displays the
corresponding mean diameters, obtained by using the saturation magnetization
$M_{\rm S}$ $\sim$ 460 emu/cm$^{3}$ for the bulk magnetite and considering the
nanoparticles as spheres. 

A careful inspection of the data in Table I indicates that the mean particle diameter $d_{\rm M} \sim$ 4.9 nm, the distribution width
$s_{\rm M} \sim 0.19$, and the corresponding standard deviation $\sigma_{\rm M} \sim$ 0.9 nm, 
obtained from the magnetization data, are very similar for all samples.  This is an expected result since the samples were prepared from the same nanoparticles starting batch. In addition, these
values are in excellent agreement with the ones obtained from TEM analysis, which were $d_{\rm T} \sim$ 5.5 nm, $s_{\rm T} \sim$ 0.15,
and  $\sigma_{\rm T} \sim$ 0.7 nm.

\begin{table}
\caption{\label{tab:table1} Mean magnetic moment ($\mu_{\rm m}$), mean magnetic diameter ($d_{\rm M}$),
distribution width ($s_{\rm M}$), and standard deviation ($\sigma_{\rm M}$) obtained from the
Langevin fits weighted by a lognormal magnetic moment distribution.}
\begin{tabular}{cccccc}
Sample & \%MM &  $\mu _{\rm m}$($\mu _{\rm B}$) & $d_{\rm M}$(nm) & $s_{\rm M}$ &  $\sigma _{\rm M}$(nm)\\
\hline D63   & 63 & 3157 & 4.8   & 0.19 & 0.9\\
O65 & 65    & 3405  & 4.9 & 0.17 &  0.9\\
O76 & 76    & 3528  & 5.0   & 0.19 & 1.0\\
O78 & 78    & 3190  & 4.8   & 0.19 & 0.9\\
\end{tabular}
\end{table}

We have also observed, when the temperature is increased, that the magnitude of $M_{\rm S}$
decreases accordingly and its saturation value is progressively
reached at higher external magnetic fields. Mostly of the $M_{\rm S}$ values
at room temperature attained up to $\sim$ 80\% of the bulk value
$\sim$ 90 emu/g. These values of $M_{\rm S}$ are quite impressive
and well above the ones usually found in the literature,\cite{GOY1, GEE1}
but in line with those obtained for NP of similar sizes.\cite {GUA1}
The high values of $M_{\rm S}$ found in our 
samples may be related to the nature of the coating used. It seems that both acids
act in order to decrease the spin disorder at the surface of the
nanoparticles, enhancing $M_{\rm S}$ values. This kind of
behavior has been also observed in Fe$_3$O$_4$ NP coated with
oleic acid.\cite{GUA1}

\begin{figure}[htp]
\centering
\includegraphics{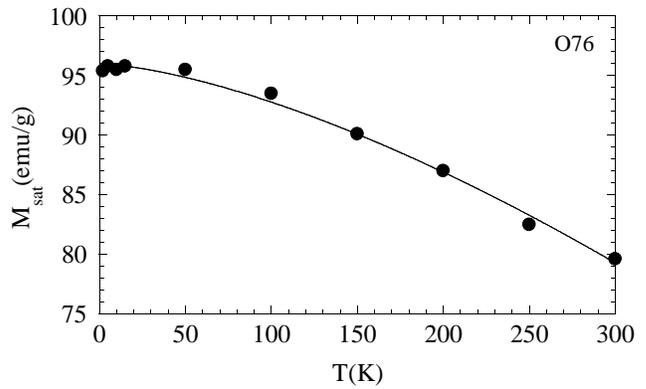}
\caption{\label{fig:epsart8} Data fitting to the Bloch's law for the sample O76.}
\end{figure}

The temperature dependence of $M_{\rm S}$ is of interest and needs to be further considered. 
For a ferromagnetic or ferrimagnetic material, the decrease of $M_{\rm S}$ from its
value at 0 K, $M_{0}$, with increasing $T$ is
related to the excitation of spin-waves with long wavelengths. For
a continuous distribution of spin-wave states, the temperature
dependence of the magnetization is expected to follow the Bloch's law 
$M_{\rm S}(T) = M_{0}(1-BT^{3/2})$, where $M_{0}$=$M_{\rm S}$($T$=0) and $B$
is a parameter proportional to the inverse of the exchange constant $J$.

A commonly observed feature in systems comprised of nanosized particles is 
a deviation from the Bloch's law at low temperatures. Such a 
deviation is primarily attributed to magnons with wavelengths larger than 
the particle dimensions that cannot be excited in nanosized materials. This
results in a gap in the energy levels of the system and the spin
waves are generated only when a threshold of thermal energy is
achieved.\cite{MAN1} In any event, we have successfully fitted our $M_{\rm S}$
data by using the Bloch's law, as displayed in Figure 8
for the sample O76. Table II shows the parameters $M_{\rm S}(0)$ and 
$B$, obtained from the fittings to the Bloch's law for all samples. The 
values of $B$, roughly ranging from $3.28$ to 3.56x$10^{-5} K^{-3/2}$,
are essentially the same when compared to the one of $B \sim 3.3$x$10^{-5}$ K$^{-3/2}$
found in NP of magnetite with similar sizes.\cite {GOY1} Table II also shows the 
corresponding values of the exchange constant $J_{AB}$ obtained from values of $B$.

In a first approximation and for magnetite, $B$ is close related to the dominant exchange 
constant $J_{AB}$ between tetrahedral (sublattice $A$) and octahedral (sublattice $B$) sites by:\cite{ARA1}

\begin{eqnarray}
B = \frac{0.05864}{4(S_{B1}+S_{B2}-S_{A})} \left[ \frac{16(S_{B1}+S_{B2}-S_{A})k_{\rm B}} {11J_{AB}S_{A}(S_{B1}+S_{B2})}\right ]^{3/2}
\end{eqnarray}

where $S_{A}=S_{B1}=5/2$ is the spin of the Fe$^{3+}$ and $S_{B2}=2$ 
the spin of the Fe$^{2+}$ ions located in the octahedral magnetic sublattice. The 
spin wave stiffness constant $D$ has been also evaluated by using the relationship 
$D=11J_{AB}S_{A}S_{B}a^{2}/(16|S_{A}-S_{B}|)$,\cite{SRI1} where
$a \sim 8.4$ \AA $ $ is the lattice parameter of the unit cell and $S_{B}=(S_{B1}+S_{B2})/2=9/4$. 

\begin{table}
\caption{\label{tab:table2} $M_{\rm S}(0)$ and $B$ parameters obtained from the
fitting to the Bloch's law, as well as the integral constant $J_{AB}$, and the spin wave stiffness constant $D$.}
\begin{tabular}{cccccc}

Sample & \% MM & $M_{\rm S}(0)$ & $B$ (x$10^{-5}$) & $J_{AB}$ & $D$\\
& & emu/g& K$^{-3/2}$ & K &meV\AA$^{2}$\\
\hline D63   & 63 & 87.4 & 3.56 & 9.0 & 100\\
O65 & 65 & 83.2 & 3.54 & 9.1 & 106\\
O76 & 76 & 96.0 & 3.35& 9.4 & 110\\
O78 & 78 & 88.5 & 3.28& 9.5 & 112\\
\end{tabular}
\end{table}

The estimated values, displayed in Table II, indicate that $B$ decreases systematically as the 
concentration of magnetic material increases, a result followed by the values of $J_{AB}$. 
The progressive increase of $J_{AB}$ with increasing volume fraction of the magnetic material
indicates that $J_{AB}$ represents an effective interaction between magnetic ions, including
the interaction of those located at the surface. The values of $J_{AB}$ $\sim 9$ K estimated
in our samples were found to be twice smaller than the one of $\sim 23$ K obtained in magnetite
single crystals by means of magnetization data.\cite{ARA1}
We also notice that the estimated values of $D \sim 100$ meV\AA$^{2}$ in our series are smaller
than the single crystal value of $\sim 320$ meV\AA$^{2}$. Such a small discrepancy is certainly
related to the fact that there is a smaller number of nearest neighbors 
for the surface spins, leading to a smaller effective $J_{AB}$ than the
one obtained in single crystals.

\begin{figure}[htp]
\centering
\includegraphics{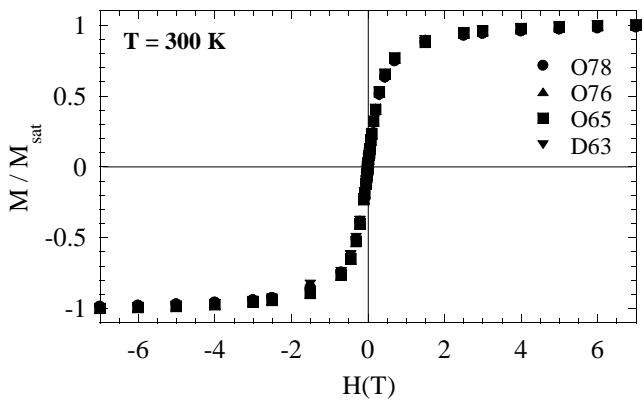}
\caption{\label{fig:epsart9} Curves of $M/M_{\rm S}$ vs. $H$ taken at $T$ = 300 K
for all samples. The curves collapse into a universal curve, as discussed in the text.}
\end{figure}

We finally notice that curves of $M/M_{\rm S}$ vs. $H$ at a given temperature, for different magnetic concentrations
and coatings also overlap, as shown in Figure 9 for $T = 300$ K. This indicates that the nature of the coating acid 
and the dipolar interactions have only a quantitative effect on the $M$ vs. $H$
curves, but the qualitative features of the NP remain the same, and are well described by the
intrinsic properties of the NP core.

\section*{Conclusions}

In summary, samples comprised of magnetite NP coated with oleic and dodecanoic acids and
with different magnetic concentrations have been synthesized and display
superparamagnetic behavior at high temperatures. The NP ensemble
may not be treated as a noninteracting monodomain system. The
dipolar interactions between NP was found to play a role,
although the overall magnetic behavior, mostly due to the same magnetic core, is preserved in 
samples with different volume fractions of the magnetic material.
The NP diameters estimated from magnetization curves were found to be comparable
with the ones obtained from TEM results, indicating that the magnetite core of the NP is the same for
all particles. The main effect related to the increase
in the magnetic concentration is the systematic increase of the blocking temperature $T_{\rm B}$ due
to an increase of the effective magnetic interaction between NP.
Therefore, changing the NP coating resulted in no appreciable changes in the overall magnetic behavior of the samples. The high values of $M_{\rm S}$, corresponding to $\sim$ 80\% of the bulk value, make these magnetite NP systems suitable for technological applications.

\begin{acknowledgments}
This work was supported by the Brazilian agency
Funda\c{c}\~{a}o de Amparo \`{a} Pesquisa do Estado de S\~{a}o
Paulo (FAPESP) under Grant No. 2005/53241-9. Three of us (R.F.J., 
F.B.E., and L.M.R.) acknowledge the Conselho Nacional de Desenvolvimento
Cient\'{i}fico e Tecnol\'{o}gico (CNPq) for fellowships.
\end{acknowledgments}


\begin{thebibliography}{10}
\bibitem{AZE1}M. Azeggagh and H. Kachkachi, Phys. Rev. B \textbf{75}, 174410 (2007).
\bibitem{NEE1}L. N\'{e}el, Ann. Phys. \textbf{3}, 137 (1948).
\bibitem{HAM1}W.C. Hamilton, Phys. Rev. \textbf{110}, 1050 (1958).
\bibitem{JAC1}See, for instance, M. J. Jacinto, P. K. Kiyohara, S. H. Masunaga, R. F. Jardim, and L. M. Rossi, Appl. Cat. A: Gen. \textbf{338}, 52 (2008); M. J. Jacinto, O. H. C. F. Santos, R. F. Jardim, R. Landers, and L. M. Rossi, Appl. Cat. A: Gen. \textbf{360}, 177 (2009).
\bibitem{TAR1}P. Tartaj, M. P. Morales, S. Veintemillas-Verdaguer, T. Gonz\'{a}lez-Carre\~{n}o, and C. J. Serna, J. Phys. D: Appl. Phys. \textbf{36}, R182 (2003).
\bibitem{ZHA1}L. Zhang, R. He, and H. C. Gu, Appl. Surf. Sci. \textbf{253}, 2611 (2006).
\bibitem{MAR1}B. Mart\'inez, X. Obradors, Ll. Balcells, A. Rouanet, and C. Monty, Phys. Rev. Lett. \textbf{80}, 181 (1998).
\bibitem{BER1}A. E. Berkowitz, R. H. Kodama, S. A. Makhlouf, F. T. Parker, F. E. Spada, E. J. McNiff  Jr., and S. Foner, J. Magn. Magn. Mater. \textbf{196}, 591 (1999).
\bibitem{CAI1}C. Caizer, Appl. Phys. A \textbf{80}, 1745 (2005).
\bibitem{GUA1}P. Guardia, B. Batlle-Brugal, A. G. Roca, O. Iglesias, M. P. Morales, C. J. Serna, A. Labarta, and X. Batlle, J. Magn. Magn. Mater. \textbf{316}, e756 (2007) and references therein.
\bibitem{SUN1}S. H. Sun and H. Zeng, J. Am. Chem. Soc. \textbf{124}, 8204 (2002).
\bibitem{SUN2}S. H. Sun, H. Zeng, D. B. Robinson, S. Raoux, P. M. Rice, S. X. Wang, and G. X. Li, J. Am. Chem. Soc. \textbf{126}, 273 (2004).
\bibitem{KAC1}H. Kachkachi, W. T. Coffey, D. S. F. Crothers, A. Ezzir, E. C. Kennedy, M. Nogu\`{e}s, and E. Tronc, J. Phys.: Condens. Matter \textbf{48}, 3077 (2000).
\bibitem{LUO1}W. Luo, S. R. Nagel, T. F. Rosenbaum, and R. E. Rosensweig, Phys. Rev. Lett. \textbf{67}, 2721 (1991).
\bibitem{DOR1}J.L. Dormann, D. Fiorani, and E. Tronc, J. Magn. Magn. Mater. \textbf{184}, 262 (1998).
\bibitem{OTE1}J. Garc\'{i}a-Otero, M. Porto, J. Rivas, and A. Bunde, Phys. Rev. Lett. \textbf{84}, 167 (2000).
\bibitem{Goya}G. F. Goya, F. C. Fonseca, R. F. Jardim, R. Muccillo, N. L. V. Carre\~{n}o, E. Longo, and E. R. Leite, J. Appl. Phys. \textbf{93}, 6531 (2003).
\bibitem{DOR3}J. L. Dormann, D. Fiorani, and E. Tronc, Adv. Chem. Phys. 98, 283 (1997).
\bibitem{JON}T. Jonsson, J. Mattsson, P. Nordblad, and P. Svedlindh, J. Magn. Magn. Mater. \textbf{168}, 269 (1997).
\bibitem{BAT1}X. Batlle and A. Labarta, J. Phys. D \textbf{35}, R15 (2002).
\bibitem{DJU} C. Djurberg, P. Svedlindh, P. Nordblad, M. F. Hansen, F. B\o dker, and S. M\o rup, Phys. Rev. Lett. \textbf{79}, 5154 (1997).
\bibitem{MYD} A. Mydosh, Spin Glasses: An Experimental Introduction (Taylor and Francis, London) 1993.
\bibitem{DOR2}J. L. Dormann, R. Cherkaoui, L. Spinu, M. Nogu\`es, F. Lucari, F. D'Orazio, D. Fiorani, A. Garcia, E. Tronc, and J.P. Jolivet, J. Magn. Magn. Mater. \textbf{187}, L139 (1998).
\bibitem{HAN2}M. F. Hansen, P. E. J\"{o}nsson, P. Nordblad, and P. Svedlindh, J. Phys.: Condens. Matter \textbf{14}, 4901 (2002).
\bibitem{SUE}S. H. Masunaga, R. F. Jardim, P. F. P. Fichtner, and J. Rivas, Phys. Rev. B \textbf{80}, 184428 (2009).
\bibitem{DOR4}J. L. Dormann, L. Bessais, and D. Fiorani, J. Phys. C \textbf{21}, 2015 (1988).
\bibitem{FON1}F. C. Fonseca, G. F. Goya, R. F. Jardim, R. Muccillo, N. L. V. Carren\~{o}, E. Longo, and E. R. Leite, Phys. Rev. B \textbf{66}, 104406 (2002).
\bibitem{GOY1}G. F. Goya, T. S. Berqu\'{o}, F. C. Fonseca, and M. P. Morales, J. App. Phys. \textbf{94}, 3520 (2003). 
\bibitem{GEE1}S. H. Gee, Y. K. Hong, D. W. Erickson, M. H. Park, and J. C. Sur, J. App. Phys. \textbf{93}, 7560 (2003). 
\bibitem{MAN1}K. Mandal, S. Mitra and P. A. Kumar, Europhys. Lett. \textbf{75}, 618 (2006).
\bibitem{SRI1}C. M. Srivastava and R. Aiyar, J. Phys. C \textbf{20}, 1119 (1987).
\bibitem{ARA1}R. Arag\'{o}n, Phys. Rev. B \textbf{46}, 5328 (1992).
\end{thebibliography}
\end{document}